\documentclass[journal,twocolumn]{IEEEtran}
\usepackage{slashbox,graphicx,times,threeparttable,amsmath,amssymb,cite,subfigure,stfloats,booktabs, url,multirow,afterpage}
\usepackage[lined,ruled]{algorithm2e}
\allowdisplaybreaks
\makeatletter

\newcommand{\Rmnum}[1]{\expandafter\@slowromancap\romannumeral #1@}
\makeatother

\begin{document}
\title{Multi-Branch Mutual-Distillation Transformer for EEG-Based Seizure Subtype Classification}

\author{Ruimin Peng, Zhenbang Du, Changming Zhao, Jingwei Luo, \\Wenzhong Liu, Xinxing~Chen, Dongrui~Wu
\thanks{R. Peng, Z. Du, W. Liu and D. Wu are with the Belt and Road Joint Laboratory on Measurement and Control Technology, Huazhong University of Science and Technology, Wuhan 430074, China.}
\thanks{C. Zhao is with the AI Platform, Software Engineering Research Center, Dongfeng Corporation Research \& Development Institute, Wuhan 430072, China. }
\thanks{J. Luo is with the China Electronic System Technology Co., Ltd., Beijing 100089, China.}
\thanks{X. Chen is with the Guangdong Provincial Key Laboratory of Human-Augmentation and Rehabilitation Robotics in Universities, Southern University of Science and Technology, Shenzhen, 518055, China.}
\thanks{X. Chen and D. Wu are the corresponding authors. E-mail: chenxx@sustech.edu.cn, drwu@hust.edu.cn.}}

\maketitle

\begin{abstract}
Cross-subject electroencephalogram (EEG) based seizure subtype classification is very important in precise epilepsy diagnostics. Deep learning is a promising solution, due to its ability to automatically extract latent patterns. However, it usually requires a large amount of training data, which may not always be available in clinical practice. This paper proposes Multi-Branch Mutual-Distillation (MBMD) Transformer for cross-subject EEG-based seizure subtype classification, which can be effectively trained from small labeled data. MBMD Transformer replaces all even-numbered encoder blocks of the vanilla Vision Transformer by our designed multi-branch encoder blocks.  A mutual-distillation strategy is proposed to transfer knowledge between the raw EEG data and its wavelets of different frequency bands. Experiments on two public EEG datasets demonstrated that our proposed MBMD Transformer outperformed several traditional machine learning and state-of-the-art deep learning approaches. To our knowledge, this is the first work on knowledge distillation for EEG-based seizure subtype classification.
\end{abstract}

\begin{IEEEkeywords}
Transformer, knowledge distillation, EEG, seizure subtype classification
\end{IEEEkeywords}

\IEEEpeerreviewmaketitle

\section{Introduction}

Epilepsy is a widespread neurological disorder characterized by the rapid and early abnormal electrical activity of neurons in the brain, affecting more than 50 million people globally\cite{lasefr2017epilepsy,wong2023eeg}. Among them, over 30\% have intractable epilepsy, which significantly impacts the patients' emotional, behavioral, and cognitive functions, severely limiting their ability to engage in daily activities\cite{singhal2023unveiling, wong2023eeg}. Furthermore, disruptions in cognition and consciousness in severe cases impose significant risks to the patient's safety and well-being. Consequently, precise diagnosis and effective treatment for epilepsy are very important.

Clinical diagnosis of epilepsy heavily relies on the expertise of medical professionals to analyze patients' electroencephalogram (EEG), which is demanding and time-consuming. Therefore, an automatic seizure diagnosis system, which analyzes EEG recordings automatically and rapidly, is highly desirable. Seizure detection, i.e., recognizing and marking the ictal fragments in EEG recordings, has been extensively studied in the literature; however, seizure subtype classification, which is critical in determining the appropriate therapies with medicine or surgery\cite{tang2022selfsupervised}, has not received enough attention.

The 2017 International League Against Epilepsy guideline \cite{scheffer2017ilae, fisher2017operational} categorizes epileptic seizures into generalized seizures [e.g., absence seizures (ABSZ), tonic seizures (TNSZ), and tonic-clonic seizures (TCSZ)], focal seizures (FSZ), and a combination of generalized and focal seizures. Different seizure types may have different prorogation patterns, e.g., ABSZ diffuses to the entire brain, whereas FSZ only affects a local area. Additionally, many seizures are caused by specific and personalized diseased tissue. These characteristics make it very challenging for automatic cross-patient seizure subtype classification.

Conventional seizure subtype classification approaches usually involve three steps: data pre-processing, feature extraction, and classification \cite{boonyakitanont2020review}. Previous studies have extracted a large number of features to be used in machine learning models, e.g., support vector machine (SVM)\cite{li2013feature}, ridge classifier (RC), logistic regression (LR)\cite{samiee2014epileptic}, and gradient boosting decision tree (GBDT). However, these manually extracted features may not be optimal.

Deep neural networks, e.g., convolutional neural networks (CNNs), recurrent neural networks, and autoencoders, have also been extensively used for automatic EEG feature extraction\cite{shoeibi2021epileptic}. Recently, Transformer\cite{vaswani2017attention} based models have achieved great success in numerous tasks.

\begin{figure}[htbp]\centering
	\subfigure[]{\label{fig:1KD}   \includegraphics[width=.8\linewidth,clip]{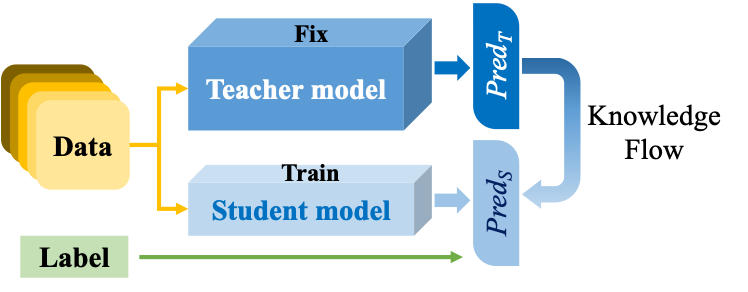}}
	\subfigure[]{\label{fig:1ML}    \includegraphics[width=.8\linewidth,clip]{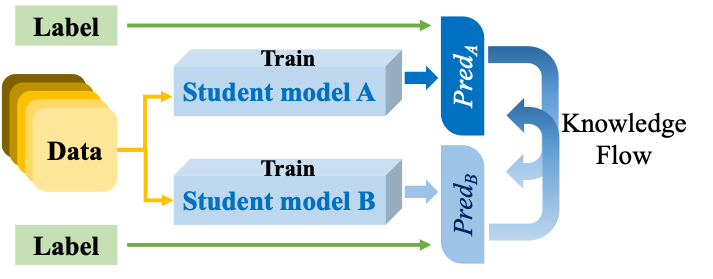}}
	\caption{(a) Knowledge distillation; and, (b) mutual learning.} \label{fig:KD}
\end{figure}

A deep learning model may be very large. To reduce the model size and enhance the training efficiency, various techniques for model compression have been proposed, e.g., pruning\cite{blalock2020state}, low-rank approximation\cite{yu2017compressing}, quantization\cite{quantization}, compact network design\cite{zhang2018shufflenet}, and knowledge distillation \cite{hinton2015distilling}. Knowledge distillation typically employs a large teacher model to provide soft labels, guiding the training of a more compact student network. By leveraging the teacher's knowledge, the smaller student model mimics the teacher's output, achieving comparable or even better performance. Different from the conventional use of a fixed teacher model, Zhang \textit{et al.}\cite{mutuallearning} proposed mutual learning, where every student model can learn from others. Fig.~\ref{fig:KD} illustrates the general ideas of knowledge distillation and mutual learning.

\begin{figure}[htbp]\centering
	\subfigure[]{\label{fig:2DA}   \includegraphics[width=.7\linewidth,clip]{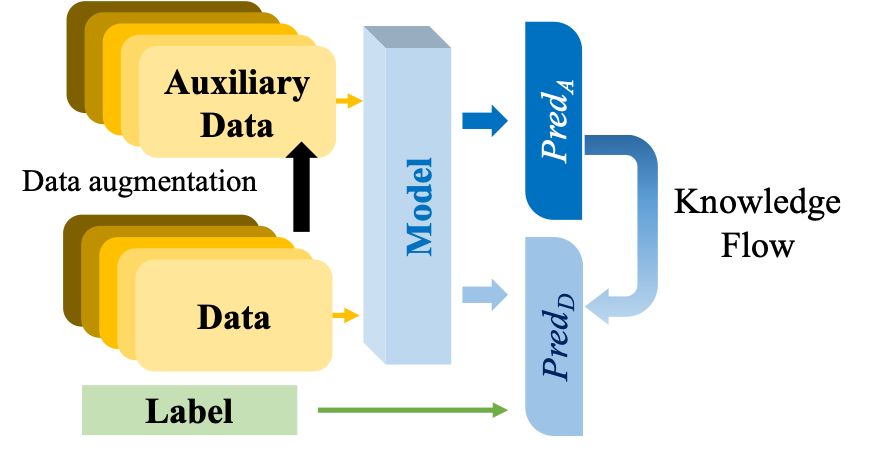}}
	\subfigure[]{\label{fig:2AS}    \includegraphics[width=.63\linewidth,clip]{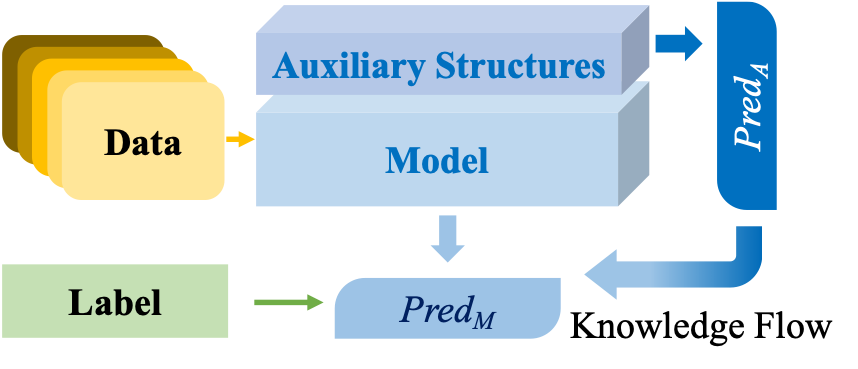}}
	\caption{Self-distillation strategies. (a) Data augmentation; and, (b) auxiliary structure.} \label{fig:self-distillation}
\end{figure}

Most model compression techniques require a large teacher model trained on big data. However, for data scarcity scenarios like seizure subtype classification, a large teacher model may not be available. In this situation, self-distillation could be used to distill a network's knowledge and guide its training. Self-distillation can be implemented by using data augmentation or auxiliary structure \cite{yang2022mixskd}, as illustrated in Fig.~\ref{fig:self-distillation}. Data augmentation based self-distillation enforces consistent predictions among augmented copies of the same instance or two instances from the same class. Auxiliary structure based self-distillation designs additional branches for the backbone network, and enforces them to be similar.

This paper proposes a novel multi-branch mutual-distillation (MBMD) Transformer for EEG-based seizure subtype classification, which can be effectively trained from small labeled data. Our main contributions are:
\begin{enumerate}
\item We design a novel multi-branch Transformer. It uses multi-branch encoder blocks, which employ a multi-branch feedforward network (FFN) to process the wavelets decomposed from the raw EEG signals with different frequency bands. All wavelets use the same class label as the raw data, and the block output is an ensemble of all branches.
\item We propose a novel mutual-distillation strategy to facilitate knowledge transfer between the raw EEG data and the wavelets. It enables the model to uncover more hidden information and achieve improved performance.
\item To our knowledge, this study is the first attempt to introduce self-distillation in cross-patient seizure subtype classification.
\end{enumerate}

The remainder of this paper is organized as follows. Section II briefly reviews related works. Section III describes the details of our proposed MBMD Transformer. Section IV presents the performance of MBMD Transformer on two seizure subtype classification datasets. Finally, Section V draws conclusions and outlines some future research directions.

\section{Related works}

This section reviews prior works on EEG-based seizure subtype classification and self-distillation.

\subsection{EEG-based Seizure Subtype Classification}

Both traditional machine learning and deep learning have been used in EEG-based seizure subtype classification.

An important consideration in traditional machine learning is to extract meaningful features from EEG signals. Vanabelle \textit{et al.}\cite{vanabelle2020epileptic} employed 22 features from the time and frequency domains to train an XGBoost model. Tian \textit{et al.} \cite{tian2019deep} used time, frequency, and time-frequency domain features in multiple classifiers for seizure detection. Zhao \textit{et al.} \cite{zhao2023source} extracted 41 time/spectral/time-frequency domain and nonlinear features to train semi-supervised and unsupervised transfer boosting algorithms for seizure subtype classification.

Many deep learning approaches have been proposed to eliminate manual feature extraction. Li \textit{et al.} \cite{li2020epileptic} proposed CE-stSENet, a multi-scale Squeeze-and-Excitation network\cite{hu2018squeeze} to extract temporal and spectral representations. Peng \textit{et al.} \cite{peng2022tie} developed a time information enhancement module to improve the classical EEGNet \cite{lawhern2018eegnet}. They further proposed Wavelet2Vec\cite{10097183}, which combined wavelet decomposition with Vision Transformer (ViT)\cite{dosovitskiy2020vit}. Tang \textit{et al.} \cite{tang2022selfsupervised} developed a self-supervised algorithm to train a recurrent graph neural network.

\subsection{Self-distillation}

Self-distillation distills knowledge from the internal network rather than external ones. An important consideration is how to acquire additional knowledge.

From the perspective of data augmentation, Xu and Liu \cite{xu2019data} proposed Data-Distortion Guided Self-Distillation (DDGSD) for training CNNs, e.g., ResNet\cite{he2016deep}. It employed random mirroring and cropping techniques to augment data and enforced consistency in predictions across different augmentations of the same image during network training. Similarly, Yun \textit{et al.} proposed class-wise self-knowledge distillation \cite{yun2020regularizing}, which randomly samples an auxiliary batch sharing the same label as the primary training batch and aligns their predictions.

Another self-distillation methodology involves the design of auxiliary structures. Zhao \textit{et al.} devised Be Your Own Teacher (BYOT) \cite{zhang2019your} that embedded classifiers for the shallow ResBlocks of ResNet. These shallow branches functioned as students to learn knowledge from the prediction provided by the deepest block, thereby enhancing the learning capacity of the shallow blocks. Lan \textit{et al.} designed an On-the-Fly Native Ensemble (ONE)\cite{zhu2018knowledge}, which added multiple branches of high-level ResBlocks on shared low-level ones. During training, the knowledge of ensemble prediction would be distilled into individual branches for enhancing model learning. Hou \textit{et al.}\cite{hou2019learning} presented a Self Attention Distillation (SAD) approach, which conducted layer-wise and top-down attention distillation to augment the representation learning process for lane detection. Ge \textit{et al.}\cite{ge2021self} proposed the BAtch Knowledge Ensembling (BAKE) technique, refining soft targets for anchor images by propagating and ensembling knowledge from other samples within the same mini-batch. Furthermore, the recently proposed Self-Knowledge-Distillation from image Mixture (MixSKD) \cite{yang2022mixskd} leveraged Mixup\cite{zhang2017mixup}, a popular data augmentation approach, with auxiliary feature alignment modules to transform feature maps from shallow layers to match with the final feature map.

\section{Methodology}

This section introduces our proposed MBMD Transformer and its training strategy. The code is available at https://github.com/rmpeng/EBE-Transformer.

\subsection{Vanilla ViT for EEG Signal Classification}

\begin{figure}[htbp]         \centering
	\includegraphics[width=\linewidth,clip]{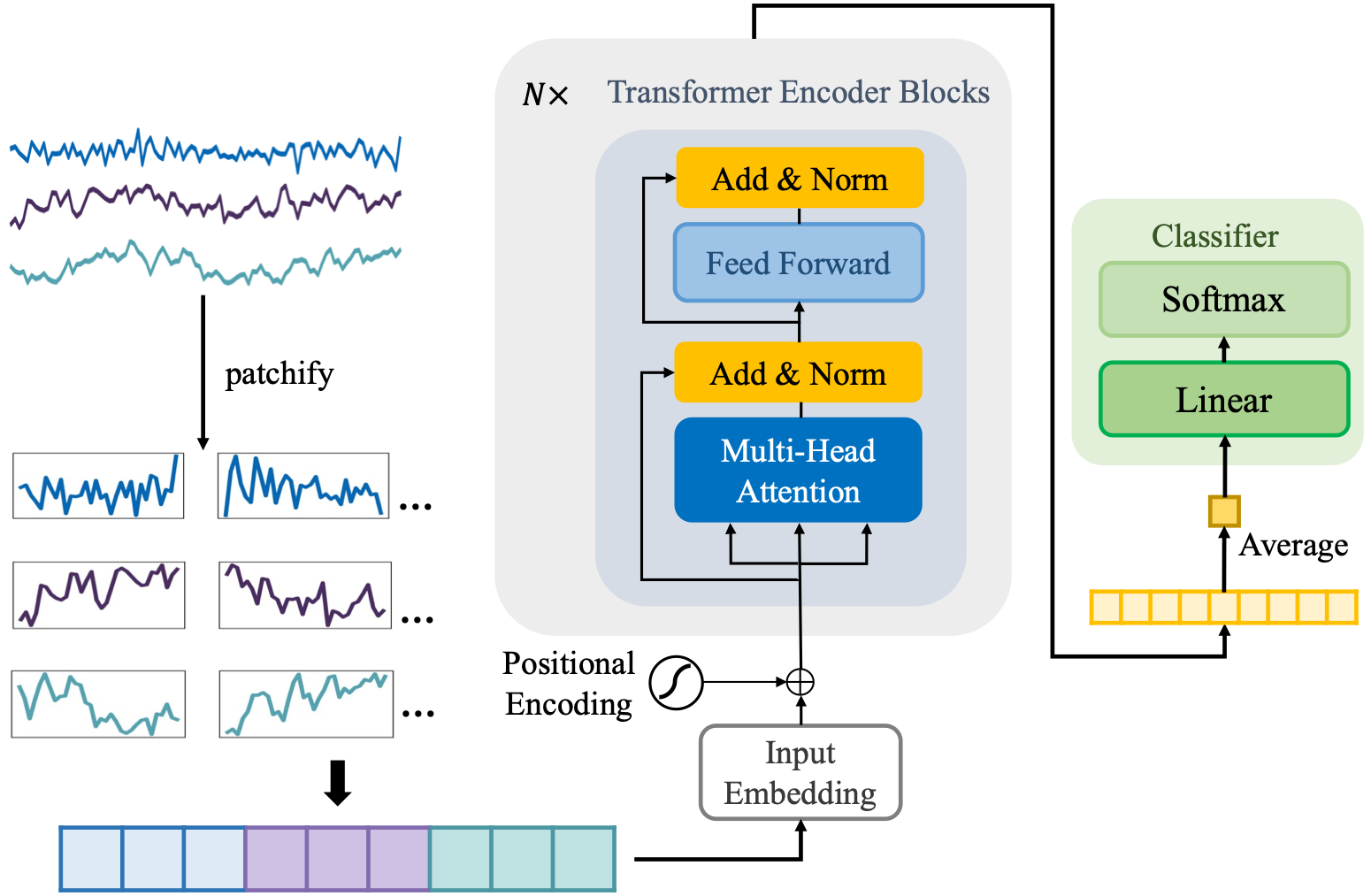}
	\caption{A vanilla ViT for EEG signal classification.} \label{fig:ViT}
\end{figure}

Fig.~\ref{fig:ViT} illustrates the training process of a vanilla ViT for EEG signal classification. Our proposed MBMD Transformer is modified from it, as introduced later in this section.

EEG signals with dimensionality [$C$ $\times$ $1$ $\times$ $L$], where $C$ is the number of channels and $L$ the number of time domain sample points, are first segmented into multiple fragments and encoded with positional encoding. The generated embeddings are then input into a Transformer encoder block,  comprised of a multi-head attention layer and an FFN layer. This encoder block is repeated $N$ times to learn the latent patch representations. The average of all patch representations after the $N$th encoder block is taken as the feature for final classification.

The classical cross-entropy loss $\mathcal{L}_{ce}$ is used in training the vanilla ViT:
\begin{align}
\mathcal{L}_{ce}=-\frac{1}{K}\sum_{k=1}^{K}{\log\big(p  \left(y = k|\boldsymbol{x},\boldsymbol{\theta}\right)\big) } \label{eq:celoss}
\end{align}
where $K$ is the number of class, $\boldsymbol{x}$ the raw input EEG trial, and $\boldsymbol{\theta}$ the model parameters.

\subsection{Our Proposed MBMD Transformer}

\begin{figure*}[htbp]  \centering
	\subfigure[]{\label{fig:4training}    \includegraphics[width=.41\linewidth,clip]{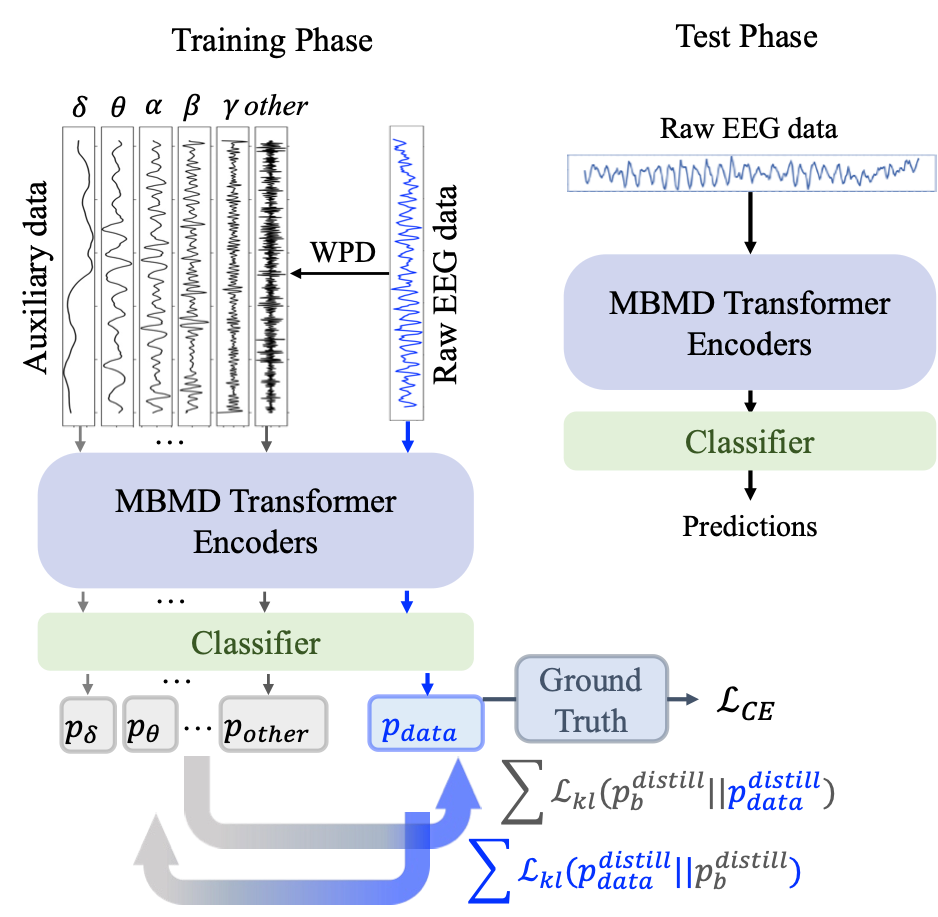}}
	\subfigure[]{\label{fig:4structure}   \includegraphics[width=.36\linewidth,clip]{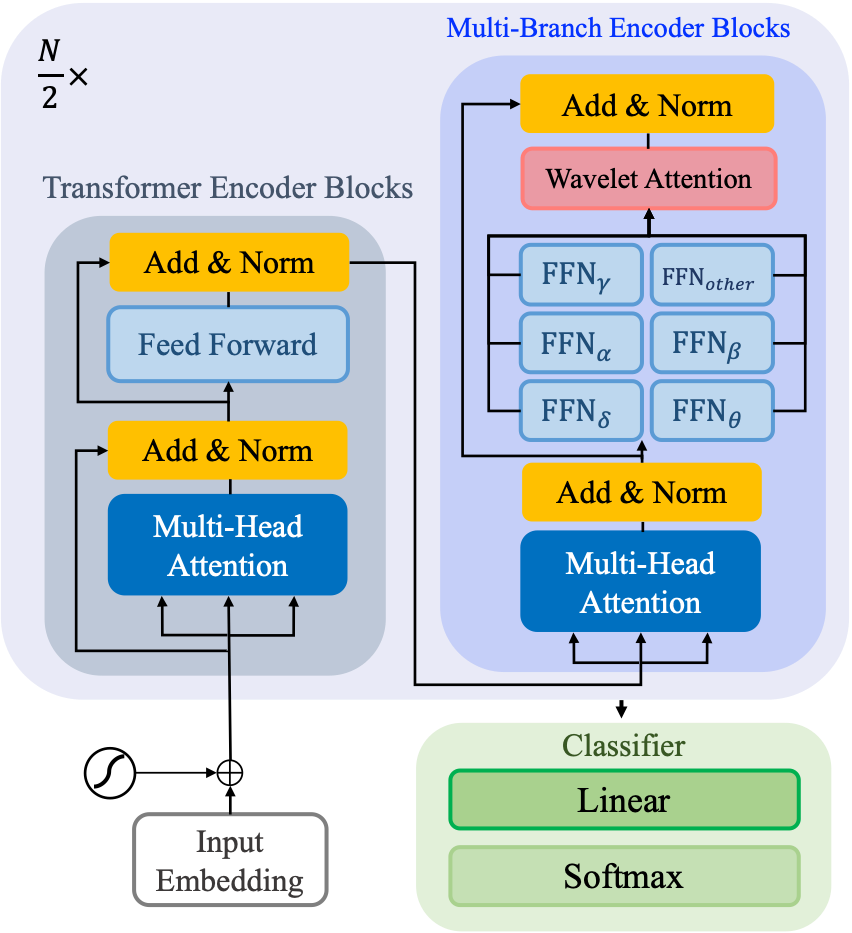}}
	\subfigure[]{\label{fig:4wavlet}    \includegraphics[width=.18\linewidth,clip]{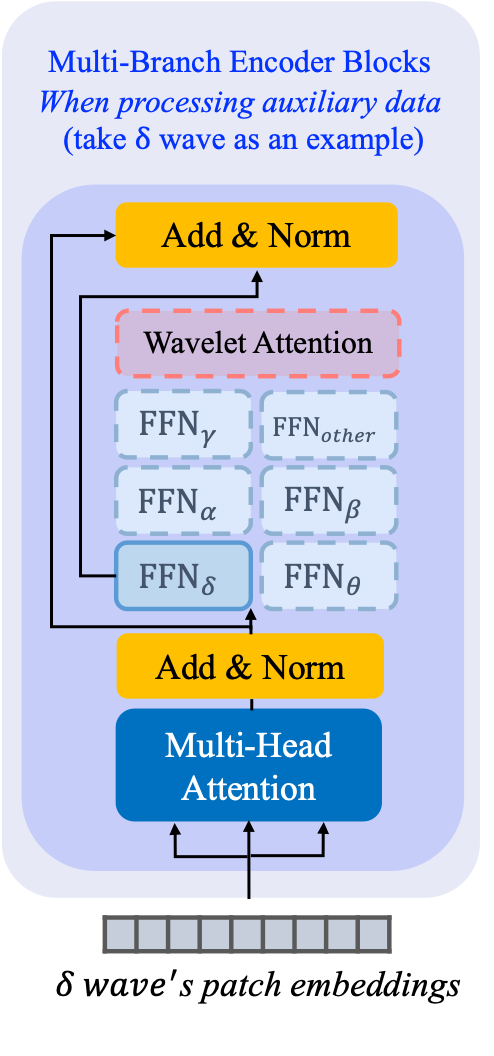}}
	\caption{MBMD Transformer with mutual-distillation. (a) Training and test; (b) the overall structure; and, (c) auxiliary data processing (take $\delta$ wave as an example).} \label{fig:Trainer}
\end{figure*}

Our proposed MBMD Transformer replaces every even-numbered encoder block in Fig.~\ref{fig:ViT} by a multi-branch encoder block, as shown in Fig.~\ref{fig:Trainer}.

\begin{figure}[htbp]  \centering
	\includegraphics[width=.98\linewidth,clip]{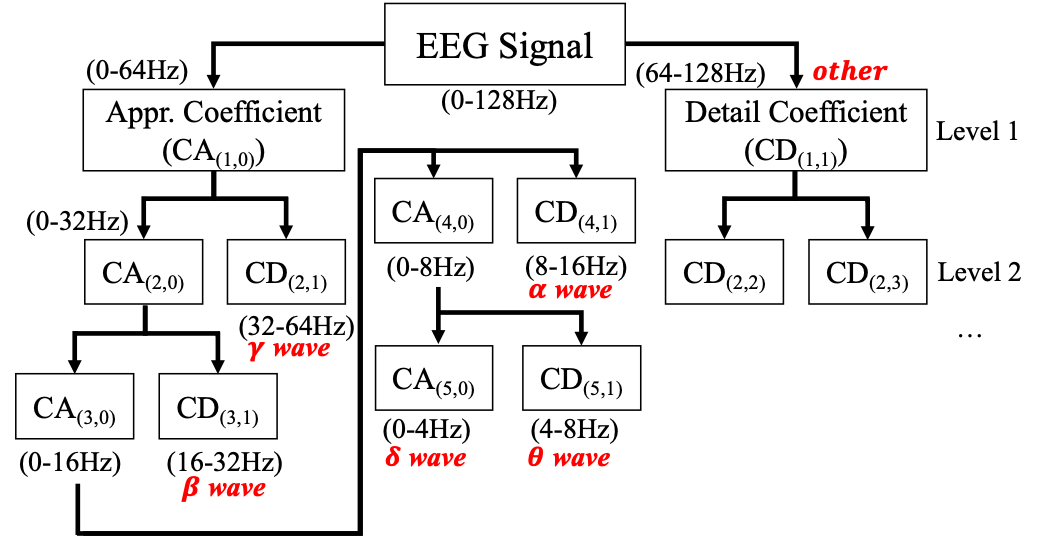}
	\caption{WPD of 128 Hz EEG signal.} \label{fig:WT}
\end{figure}

Fig.~\ref{fig:4training} explains the training and test process of MBMD Transformer. In the training phase, MBMD Transformer first uses Wavelet Packet Decomposition (WPD) \cite{10097183} to generate auxiliary wavelets corresponding to 6 different EEG frequency bands, namely, $\delta$ ($0$-$4$ Hz), $\theta$ ($4$-$8$ Hz), $\alpha$ ($8$-$16$ Hz), $\beta$ ($16$-$32$ Hz), $\gamma$ ($32$-$64$ Hz), and the remaining ($other$), which have the same label as the raw EEG trial. Fig.~\ref{fig:WT} illustrates the process of decomposing a 128 Hz EEG trial.

Next,  the raw EEG and its auxiliary wavelets enter the same linear projection layer to generate embeddings, which are fed into a traditional Transformer encoder block and our proposed multi-branch encoder block, as shown in Fig.~\ref{fig:4structure}. Different embeddings share the same multi-head attention layer and the same FFN layer in the traditional encoder block, but the same multi-head attention layer and separate FNN layers in the multi-branch encoder block (e.g., FFN$_\delta$ for $\delta$ wave, as shown in Fig.~\ref{fig:4wavlet}). Note that in the multi-branch encoder block, the raw EEG data are processed by all 6 Expert FFNs and their average is computed. A wavelet attention mechanism is developed to weight the 6 Expert FNNs, as introduced in Subsection~\ref{sect:self-distillation}. The traditional encoder block and multi-branch encoder block pair is repeated $\frac{N}{2}$ times, resulting in a total of $N$ encoder blocks. The final representations of the raw EEG data and the 6 wavelets are concatenated for classification. Predictions from the 6 wavelets also serve as soft labels for mutual-distillation, as explained in Subsection~\ref{sect:MBMD}.

In the test phase, only the raw EEG trials (but not the wavelets) are fed into the MBMD Transformer for classification. Every multi-branch encoder block sends the ensemble outputs of all 6 Expert FFNs to the next traditional encoder block.

\subsection{Mutual-Distillation} \label{sect:self-distillation}

Allen-Zhu and Li \cite{allen2020towards} introduced a `multi-view' concept to explain why ensemble/knowledge distillation succeeds in deep learning, i.e., self-distillation could be regarded as implicitly combining ensemble and knowledge distillation to improve the test accuracy. Inspired by the `multi-view' theory, we propose a mutual-distillation strategy to enhance model learning from various auxiliary wavelets. It takes the prediction from the raw EEG data, and these more accurate predictions from the auxiliary branches, as \textit{peer teachers}. Multiple peer teachers enable the model to learn more comprehensive patterns, enhancing the overall training effectiveness.

More specifically, let $\mathcal{F}_{branch}=\left\lbrace \mathcal{F}_\delta, \mathcal{F}_{\theta}, \mathcal{F}_{\alpha}, \mathcal{F}_{\beta}, \mathcal{F}_{\gamma}, \mathcal{F}_{other}\right\rbrace$ be the feature sets learned by models $\mathcal{M}_{branch}=\left\lbrace \mathcal{M}_\delta, \mathcal{M}_{\theta}, \mathcal{M}_{\alpha}, \mathcal{M}_{\beta}, \mathcal{M}_{\gamma}, \mathcal{M}_{other}\right\rbrace$ with their corresponding Expert FFNs, and $\mathcal{F}_{data}$ be the feature set of raw data learned by the ensemble model $\mathcal{M}_{data}$. Mutual-distillation aligns $\mathcal{M}_{data}$ and $\mathcal{M}_{branch}$, so they together can learn a larger feature set $\mathcal{F}_{data}\cup\mathcal{F}_{branch}$.

As illustrated in the training phase of Fig.~\ref{fig:4training}, knowledge from the raw EEG data is transferred to supervise the training of each branch, whereas the predictions from the branches are also used to improve the performance on the raw data.

Generally, knowledge distillation adopts Kullback-Leibler divergence $\mathcal{L}_{kl}$ to measure the consistency between the prediction probability distributions of the student model $\boldsymbol{p_{s}}$ and the teacher model $\boldsymbol{p_{t}}$ \cite{hinton2015distilling}:
\begin{equation} \label{eq:kl}
\begin{aligned}
&\mathcal{L}_{kl}(\boldsymbol{p_{t}}||\boldsymbol{p_{s}})
=  \\
& -\frac{1}{K}
\sum_{k=1}^{K}p_{t}\left(y = k|\boldsymbol{x},\boldsymbol{\theta}\right) log \frac{p_{t}\left(y = k|\boldsymbol{x},\boldsymbol{\theta}\right)}{p_{s}\left(y = k|\boldsymbol{x},\boldsymbol{\theta}\right)},
\end{aligned}
\end{equation}
where $p \left(y = k|\boldsymbol{x},\boldsymbol{\theta}\right) = softmax(z^{y= k})$ is the prediction probability for class $k$. To avoid ignoring low probabilities, we introduce a distillation temperature $T$ to enhance knowledge transfer:
\begin{align}
p^{distill}\left(y = k|\boldsymbol{x},\boldsymbol{\theta}\right)=
\frac{\mathrm{exp}({{z^{y= k}}/{T}})}
{\sum^{K}_{m=1}
	{\mathrm{exp}({z^{y=m}}/{T})}}. \label{eq:pdistil}
\end{align}

To summarize, the loss for mutual-distillation $\mathcal{L}_{distill}$ is:
\begin{equation}\label{eq:lossdistil}
\begin{aligned}
&\mathcal{L}_{distill} =  \\
& \frac{1}{2} \sum^{B}_{b=1}\big(\mathcal{L}_{kl}(\boldsymbol{p}_{data}^{distill}||\boldsymbol{p}_{b}^{distill}) + \mathcal{L}_{kl}(\boldsymbol{p}_{b}^{distill}||\boldsymbol{p}_{data}^{distill})\big).
\end{aligned}
\end{equation}

\subsection{Multi-Branch Encoder Block} \label{sect:MBMD}

\begin{figure}[htbp]\centering
	\subfigure[]{\label{fig:5MBMD_b}   \includegraphics[width=\linewidth,clip]{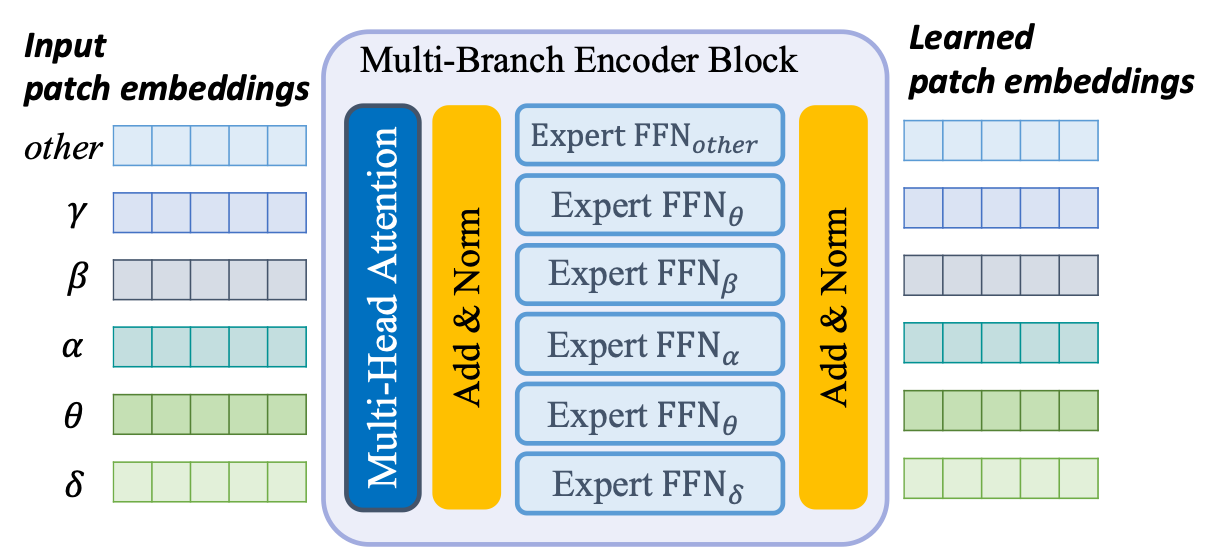}}
	\subfigure[]{\label{fig:5MBMD_d}    \includegraphics[width=\linewidth,clip]{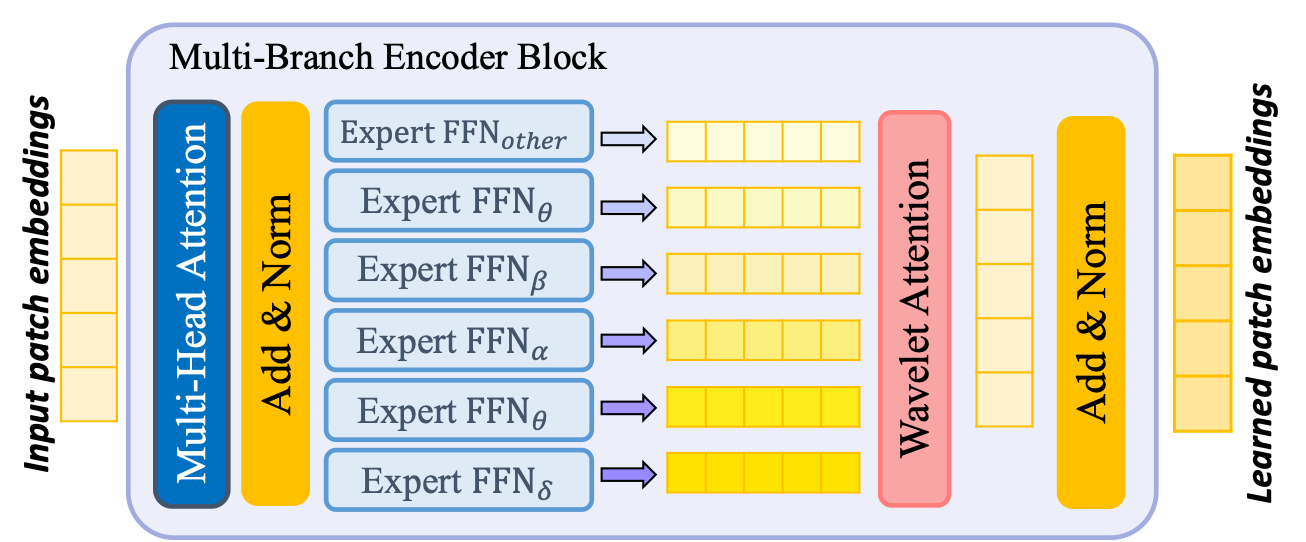}}
	\caption{The multi-branch encoder block for processing (a) the auxiliary wavelets; and, (b) the raw EEG data.} \label{fig:MBMD}
\end{figure}

The multi-branch encoder block splits the FFN layer in the traditional Transformer encoder block into multiple branches, each for a distinct wavelet frequency band. Fig.~\ref{fig:MBMD} illustrates the process.

As depicted in Fig.~\ref{fig:5MBMD_b}, each Expert FFN handles a different wavelet, ensuring the learned embeddings from different wavelets are independent, which are essential to mutual-distillation. As described in Fig.~\ref{fig:5MBMD_d}, all Expert FFNs process the raw EEG data, and the final embedding is their average.

We use a branch-wise wavelet attention mechanism to learn an adaptive weight vector $\boldsymbol{w}$ for the expert FNNs. Initially, all branches are assigned equal weights, which are iteratively adjusted during training. Before the ensemble operation, a $\mathrm{softmax}$ function is applied to normalize $\boldsymbol{w}$. The classification for raw data $\boldsymbol{z}_{data}$ is:
\begin{align}
\boldsymbol{z}_{data}= \sum_{b=1}^{B}{\mathrm{softmax}(w_b) \cdot \boldsymbol{z}_b}, \label{eq:ensemble}
\end{align}
where $B$ is the number of branches, and $w_b\in \boldsymbol{w}=\left[w_1,w_2,...,w_B \right]$ and $\boldsymbol{z}_b=\left[z^{y=1}, z^{y=2},..., z^{y=K}\right]$ are the $b$-th branch's weight and classification, respectively.

L1 regularization $\mathcal{L}_{norm}$ is used to promote the weight sparsity:
\begin{align}
\mathcal{L}_{norm}= \sum_{b=1}^{B} \left| w_b \right|  \label{eq:l1norm}
\end{align}

\subsection{Overall Loss Function for MBMD Transformer}

The overall loss function for MBMD Transformer training is:
\begin{align}
\mathcal{L}= \mathcal{L}_{ce} + \mathcal{L}_{distill} + \lambda \mathcal{L}_{norm}, \label{eq:loss}
\end{align}
where $\lambda$ is a hyperparameter to trade-off the strength of normalization, which was set to $0.01$ in our experiments.

\section{Experiments}

This section presents the experimental results on two seizure subtype classification datasets, to validate the performance of our proposed MBMD Transformer.

\subsection{Datasets, Preprocessing, and Experimental Settings}

Two public seizure datasets, CHSZ\cite{peng2022tie} and TUSZ (V1.5.2)\cite{shah2018temple}, were used. The former includes EEG recordings from 27 pediatric patients, and the latter from $68$ patients spanning all age groups. This study focused on four typical seizure subtypes: ABSZ, FSZ, TNSZ, and TCSZ. Table~\ref{tab: data} summarizes the characteristics of the two datasets.

\begin{table}[!htbp] \centering
	\setlength{\tabcolsep}{5mm}
	\caption{Summary of CHSZ and TUSZ datasets.}   \label{tab: data}
	\renewcommand\arraystretch{1.2}
	\scalebox{1}{
		\begin{threeparttable} 	\begin{tabular}{c|cccc}
				\toprule
				Dataset& ABSZ  & FSZ & TNSZ & TCSZ \\
				\midrule
				CHSZ     & 81   & 87  & 15   & 16   \\
				TUSZ     & 76   & 418 & 62   & 48  \\
				\bottomrule
			\end{tabular}
	\end{threeparttable}}
\end{table}

All recordings were first down-sampled to $128$ Hz. We then applied a $50$ Hz notch filter, a $64$ Hz low-pass filter, and detrending, to remove EEG artifacts. Next, we performed re-referencing to generate standardized $20$-channel recordings \cite{li2020epileptic}, which were then segmented using a $4$-second sliding window, with a $50\%$ overlap between two successive windows.

All deep models used batch size $32$, AdamW optimizer with weight decay $5e{-5}$, learning rate $0.001$, and early stopping with patience $10$ in training. All Transformer-based models used $4$ encoder blocks, patch size $64$, and embedding dimensionality $128$. All self-distillation approaches used identical distillation temperature $T=6$.

To consider class-imbalance, the raw accuracy (ACC), balanced classification accuracy \cite{bca} (BCA; the average of per-class accuracies), and weighted $F_1$ score\cite{f1} (the weighted average of per-class $F_1$ scores) were used together for model performance evaluation.

We followed \cite{peng2022tie} to conduct three-fold cross-patient validations. All reported results were the average of ten repeats.

\subsection{Overall Performance}

MBMD Transformer was first compared with nine existing EEG-based seizure subtype classification approaches, including four traditional approaches (SVM, RC, LR, and GBDT) and five state-of-the-art deep models (EEGNet, TIE-EEGNet, CE-stSENet, ViT, and WaveletTransformer). We followed \cite{zhao2023source} to extract $41$ features for the four traditional approaches.

Table~\ref{tab:seizure_main} demonstrates MBMD Transformer's superior performance: it obtained the highest ACC, BCA and weighted $F_1$ scores on both datasets. Particularly, compared with the 4-layer WaveletTransformer, an ensemble model of six four-layer vanilla ViTs (one for each wavelet, a total of 4*6 encoders), MBMD Transformer improved the BCA by at least $1.4\%$, using only two traditional encoder blocks and two multi-branch encoder blocks (4 encoders). When WaveletTransformer used only one layer (a total of 1*6 encoders, similar to MBMD Transformer), its performance was much worse than MBMD Transformer, suggesting that MBMD Transformer can use the encoders more effectively.

\begin{table*}[!] \centering \setlength{\tabcolsep}{5mm}
	\scriptsize
	\caption{Performance (mean$\pm$std) of different seizure subtype classification approaches on CHSZ and TUSZ datasets. The best performance in each column is marked in bold.}   \label{tab:seizure_main}
	\renewcommand\arraystretch{1}
	\scalebox{1}{
		\begin{threeparttable}
			\begin{tabular}{c|ccc|ccc}
				\toprule
				& \multicolumn{3}{c|}{CHSZ}                                                            & \multicolumn{3}{c}{TUSZ}                                                            \\
				\midrule
				\multicolumn{1}{c|}{\textbf{Approach}}& \multicolumn{1}{c}{ACC} & \multicolumn{1}{c}{BCA} & \multicolumn{1}{c|}{Weighted $F_1$} & \multicolumn{1}{c}{ACC} & \multicolumn{1}{c}{BCA} & \multicolumn{1}{c}{Weighted $F_1$} \\
				\midrule
SVM     & 0.526 ${{ \pm 0.152}}$    & 0.461 ${{ \pm  0.083}}$      & 0.498 ${{ \pm 0.109}}$     & 0.627 ${{ \pm 0.077}}$     & 0.532 ${{ \pm 0.111}}$     & 0.641 ${{ \pm  0.068 }}$    \\
RC        & 0.516 ${{ \pm 0.209}}$    & 0.474 ${{ \pm0.066 }}$      & 0.475 ${{ \pm0.183 }}$     & 0.646 ${{ \pm0.145  }}$    & 0.512 ${{ \pm 0.028 }}$     & 0.652 ${{ \pm 0.143 }}$    \\
LR        & 0.471 ${{ \pm 0.231 }}$   & 0.449 ${{ \pm 0.084  }}$     & 0.450 ${{ \pm 0.199 }}$    & 0.663 ${{ \pm 0.135 }}$    & 0.512 ${{ \pm 0.035 }}$     & 0.663 ${{ \pm0.125 }}$     \\
GBDT   & 0.621 ${{ \pm 0.113}}$    & 0.562 ${{ \pm 0.025}}$      & 0.606 ${{ \pm 0.067}}$     & 0.718 ${{ \pm 0.093}}$      & 0.459 ${{ \pm 0.077}}$      & 0.704 ${{ \pm 0.088}}$     \\
EEGNet & 0.309 ${{ \pm 0.052}}$      & 0.356 ${{ \pm 0.069}}$      & 0.309 ${{ \pm 0.039}}$     & 0.471 ${{ \pm 0.062}}$      & 0.514 ${{ \pm 0.043}}$      & 0.510 ${{ \pm 0.071}}$    \\
TIE-EEGNet & 0.593 ${{ \pm 0.034}}$      & 0.575 ${{ \pm 0.036}}$      & 0.615 ${{ \pm 0.032}}$     & 0.635 ${{ \pm 0.025}}$      & 0.561 ${{ \pm 0.027}}$      & 0.655 ${{ \pm 0.019}}$     \\
CE-stSENet  & 0.577 ${{ \pm 0.026}}$      & 0.567 ${{ \pm 0.043}}$     & 0.539 ${{ \pm 0.022}}$      & 0.745 ${{ \pm 0.070}}$      & 0.545 ${{ \pm 0.028}}$      & 0.703 ${{ \pm 0.050}}$ 	\\
ViT      & 0.539 ${{ \pm 0.067}}$        & 0.610 ${{ \pm 0.041}}$        & 0.548 ${{ \pm 0.079}}$       & 0.704 ${{ \pm 0.042}}$        & 0.657 ${{ \pm 0.064}}$        & 0.708 ${{ \pm0.045}}$       \\
WaveletTransformer (4-layer) & 0.632 ${{ \pm 0.073}}$        & 0.670 ${{ \pm 0.055}}$        & 0.642 ${{ \pm  0.077}}$    & 0.720 ${{ \pm 0.025}}$    & 0.628 ${{ \pm 0.044}}$    & 0.716 ${{ \pm 0.027}}$     \\
WaveletTransformer (1-layer) & 0.489 ${{ \pm 0.066}}$        & 0.545 ${{ \pm 0.064}}$        & 0.494 ${{ \pm  0.079}}$ & 0.718 ${{ \pm 0.025}}$    & 0.623 ${{ \pm 0.050}}$    & 0.716 ${{ \pm 0.028}}$     \\
\midrule
MBMD Transformer  & \textbf{0.650} ${{ \pm0.071}}$   & \textbf{0.684} ${{ \pm 0.066}}$   & \textbf{0.667} ${{ \pm 0.080}}$  & \textbf{0.746} ${{ \pm 0.024}}$      & \textbf{0.666} ${{ \pm 0.051}}$ & \textbf{0.739} ${{ \pm 0.030}}$     \\
				\bottomrule
			\end{tabular}
	\end{threeparttable}}
\end{table*}

\subsection{Effectiveness of Mutual-Distillation}

To demonstrate the effectiveness of mutual-distillation in MBMD Transformer, we compared it with five existing self-distillation approaches (BYOT, DDGSD, ONE, BAKE, and SAD). They were originally proposed for computer vision tasks, using CNN-based backbones. For fair comparison, we replaced their CNN-based backbones with the ViT backbone.

Table~\ref{tab:sd_main} shows the performance. On the CHSZ dataset, MBMD Transformer achieved the highest ACC and $F_1$ score, and the second-highest BCA, only lower than DDGSD by $0.005$. On the TUSZ dataset, MBMD Transformer ranked first on $F_1$ score and second on ACC and BCA (the latter two were only lower than the best by $0.001$). Overall, MBMD Transformer had the best performance.

\begin{table*}[!] \centering \setlength{\tabcolsep}{2.4mm}
	\scriptsize
	\caption{Performance (mean$\pm$std) of different distillation approaches on CHSZ and TUSZ datasets. The best performance in each column is marked in bold, and the second best with an underline.}   \label{tab:sd_main}
	\renewcommand\arraystretch{1}
	\scalebox{1}{
		\begin{threeparttable}
			\begin{tabular}{c|ccc|ccc|ccc}
				\toprule
				& \multicolumn{3}{c|}{CHSZ}                                                            & \multicolumn{3}{c|}{TUSZ}
				&	\multicolumn{3}{c}{Average Rank}                                                          \\
				\midrule
				\multicolumn{1}{c|}{\textbf{Approach}}& \multicolumn{1}{c}{ACC} & \multicolumn{1}{c}{BCA} & \multicolumn{1}{c|}{Weighted $F_1$} & \multicolumn{1}{c}{ACC} & \multicolumn{1}{c}{BCA} & \multicolumn{1}{c|}{Weighted $F_1$} & \multicolumn{1}{c}{ACC} & \multicolumn{1}{c}{BCA} & \multicolumn{1}{c}{Weighted $F_1$} \\
				\midrule
				Backbone (ViT) & 0.539 ${{ \pm 0.067}}$        & 0.610 ${{ \pm 0.041}}$        & 0.548 ${{ \pm 0.079}}$       & 0.704 ${{ \pm 0.042}}$        & 0.657 ${{ \pm 0.064}}$        & 0.708 ${{ \pm0.045}}$   & & &   \\
				\midrule
				BYOT  & 0.485 ${{ \pm 0.101}}$   & 0.569 ${{ \pm 0.048}}$   & 0.484 ${{ \pm 0.115}}$ & 0.667 ${{ \pm 0.058}}$  & 0.649 ${{ \pm 0.037}}$   & 0.679 ${{ \pm 0.063 }}$ & 5.500 & 5.500 & 5.500          \\
				DDGSD & \underline{0.624} ${{ \pm 0.093 }}$ & \textbf{0.689} ${{ \pm 0.050  }}$  & \underline{0.648} ${{ \pm 0.087 }}$  & 0.650 ${{ \pm 0.045 }}$  & 0.631 ${{ \pm 0.034 }}$  & 0.670 ${{ \pm 0.053 }}$   & 4.000           & 3.500           & 4.000          \\
				ONE   & 0.519 ${{ \pm 0.056 }}$  & 0.604 ${{ \pm 0.041 }}$    & 0.524 ${{ \pm 0.062 }}$   & 0.738 ${{ \pm 0.030 }}$   & \textbf{0.667} ${{ \pm 0.050 }}$      & 0.733 ${{ \pm 0.031 }}$     & 3.500           & 3.000           & 4.000          \\
				BAKE  & 0.621 ${{ \pm 0.062 }}$    & 0.628 ${{ \pm 0.057 }}$     & 0.628 ${{ \pm 0.066  }}$    & \textbf{0.747} ${{ \pm 0.028  }}$    & 0.658 ${{ \pm 0.057 }}$      & \underline{0.738} ${{ \pm 0.031 }}$    & 2.000           & 3.500           & 2.500          \\
				SAD  & 0.510 ${{ \pm 0.096}}$   & 0.615 ${{ \pm 0.074}}$   & 0.536 ${{ \pm 0.105 }}$    & 0.671 ${{ \pm 0.025}}$    & 0.658 ${{ \pm 0.040}}$   & 0.689 ${{ \pm 0.021}}$  & 4.500           & 3.500           & 4.000          \\ \midrule
				MBMD Transformer  & \textbf{0.650}  ${{ \pm 0.071}}$       & \underline{0.684}  ${{ \pm  0.066}}$       & \textbf{0.667}  ${{ \pm 0.080}}$      & \underline{0.746}  ${{ \pm 0.024}}$       & \underline{0.666}  ${{ \pm 0.051}}$       & \textbf{0.739} ${{ \pm 0.030 }}$     & \textbf{1.500}  & \textbf{2.000}  & \textbf{1.000}\\
				\bottomrule
			\end{tabular}
	\end{threeparttable}}
\end{table*}

Ablation studies were performed to further investigate the effectiveness of mutual-distillation. We replaced the loss item $\mathcal{L}_{distill} $ with $\mathcal{L}_{kl}^{e}$, which retains only the knowledge flow from each wavelet branch to the raw data, but not the opposite:
 \begin{equation}\label{eq:losskl_distill}
\begin{aligned}
\mathcal{L}_{kl}^{e} = \sum^{B}_{b=1} \mathcal{L}_{kl}(\boldsymbol{p}_{b}^{distill}||\boldsymbol{p}_{data}^{distill}).
\end{aligned}
\end{equation}

Table~\ref{tab:abl_sd} shows the results. Compared with the performance using $\mathcal{L}_{ce}$ only, adding $\mathcal{L}_{kl}^{e}$ always improved the performance, and replacing $\mathcal{L}_{kl}^{e}$ by our proposed $\mathcal{L}_{distill}$ further improved the performance on both datasets and for all three measures. This demonstrated the advantage of bi-directional knowledge transfer over single-directional transfer.

\begin{table*}[!] \centering \setlength{\tabcolsep}{5mm}
	\caption{Performance (mean$\pm$std) of different distillation strategies on CHSZ and TUSZ datasets. The best performance in each row is marked in bold.}   \label{tab:abl_sd}
	\renewcommand\arraystretch{1}
	\scalebox{1}{
		\begin{threeparttable}
\begin{tabular}{l|ccc|ccc}
					\toprule
	\multicolumn{1}{c|}{\multirow{2}{*}{Loss   Type}} & \multicolumn{3}{c|}{CHSZ}    & \multicolumn{3}{c}{TUSZ}        \\
					\cline{2-7}\specialrule{0em}{1pt}{1pt}
	\multicolumn{1}{c|}{}  & \multicolumn{1}{c}{ACC} & \multicolumn{1}{c}{BCA} & \multicolumn{1}{c|}{Weighted $F_1$} & \multicolumn{1}{c}{ACC} & \multicolumn{1}{c}{BCA} & \multicolumn{1}{c}{Weighted $F_1$} \\
					\midrule
	$\mathcal{L}_{ce}$   &  0.547 ${_{ \pm 0.086 }}$    & 0.621 ${_{ \pm 0.058}}$      & 0.555 ${_{ \pm 0.099}}$
	& 0.720 ${_{ \pm 0.050}}$       & 0.653 ${_{ \pm 0.057}}$         & 0.719 ${_{ \pm 0.049}}$   \\
	$\mathcal{L}_{ce} + \mathcal{L}_{kl}^{e} $  & 0.629 ${_{ \pm 0.063 }}$       & 0.648 ${_{ \pm  0.049 }}$       & 0.648 ${_{ \pm  0.080 }}$      & 0.737 ${_{ \pm 0.032}}$  & 0.658 ${_{ \pm 0.050}}$      & 0.731 ${_{ \pm 0.036}}$                              \\
	$\mathcal{L}_{ce} + \mathcal{L}_{distill} $ & \textbf{0.650} ${_{ \pm  0.071 }}$   & \textbf{0.684} ${_{ \pm 0.066}}$ & \textbf{0.667} ${_{ \pm  0.080}}$  & \textbf{0.746} ${_{ \pm 0.024}}$  & \textbf{0.666} ${_{ \pm 0.051}}$  & \textbf{0.739} ${_{ \pm 0.030}}$ \\
			\bottomrule
\end{tabular}
	\end{threeparttable}
}
\end{table*}

\subsection{Effectiveness of Wavelet Attention}

Ablation studies were also performed to investigate the effectiveness of our proposed branch-wise wavelet attention mechanism. It was compared with two baselines: an average strategy and a sample-wise gate network. The former averaged the six branches' outputs as the prediction, whereas the latter added a one-hidden-layer perceptron to predict the branch weights for each sample, inspired by Mixture-of-Experts models \cite{subasi2007eeg, ubeyli2008wavelet}. Additionally, a normalization term $L_{imp}$ \cite{shazeer2017outrageously} was incorporated to encourage all branches to have similar importance.

Table~\ref{tab:abl_wa} shows that our proposed wavelet attention mechanism with $\mathcal{L}_{norm}$ achieved the best ACC and $F_1$ score on both datasets, and also the best BCA on CHSZ. The average strategy had the best BCA on TUSZ, only $0.005$ higher than the wavelet attention mechanism. The gate network had the lowest overall performance.

\begin{table*}[!] \centering \setlength{\tabcolsep}{6mm}
	\scriptsize
	\caption{Performance (mean$\pm$std) of different ensemble strategies on CHSZ and TUSZ datasets. The best performance in each row is marked in bold.}   \label{tab:abl_wa}
	\renewcommand\arraystretch{1}
	\scalebox{1}{
		\begin{threeparttable}
			\begin{tabular}{c|c|c|cc|cc}
				\toprule
				\multirow{2}{*}{Datasets} & \multirow{2}{*}{}     & \multirow{2}{*}{Average} &\multicolumn{2}{c|}{Gate Network} & \multicolumn{2}{c}{Wavelet  Attention}\\
				& & &(w/o $\mathcal{L}_{Imp.}$)& (w/ $\mathcal{L}_{Imp.}$)&(w/o $\mathcal{L}_{norm}$)& (w/ $\mathcal{L}_{norm}$)\\
				\midrule
				\multirow{3}{*}{CHSZ} & ACC    & 0.641 ${_{ \pm 0.064}}$& 0.626 ${_{ \pm0.081 }}$& 0.627 ${_{ \pm 0.082}}$ & 0.610 ${_{ \pm 0.048}}$   & \textbf{0.650} ${_{ \pm0.071}}$    \\
				& BCA     & 0.677 ${_{ \pm0.058 }}$&  0.649 ${_{ \pm0.074 }}$& 0.664 ${_{ \pm 0.074}}$  & 0.660 ${_{ \pm0.049}}$     & \textbf{0.684} ${_{ \pm 0.066}}$     \\
				& Weighted $F_1$  & 0.662 ${_{ \pm 0.072 }}$&  0.630 ${_{ \pm0.093 }}$& 0.632 ${_{ \pm 0.096}}$  & 0.630 ${_{ \pm0.063}}$    & \textbf{0.667} ${_{ \pm 0.080}}$    \\
				\midrule
				\multirow{3}{*}{TUSZ} & ACC   & 0.740 ${_{ \pm 0.021}}$& 0.738 ${_{ \pm0.033 }}$ &0.735 ${_{ \pm0.024 }}$& 0.745 ${_{ \pm0.030 }}$  & \textbf{0.746}  ${_{ \pm 0.024  }}$    \\
				& BCA    & \textbf{0.671} ${_{ \pm0.046}}$ & 0.652 ${_{ \pm0.060 }}$ &0.660 ${_{ \pm0.056 }}$   & 0.668 ${_{ \pm 0.038  }}$     & 0.666 ${_{ \pm0.051 }}$      \\
				& Weighted $F_1$     & 0.734 ${_{ \pm0.028}}$ &0.732 ${_{ \pm0.038 }}$ & 0.730  ${_{ \pm0.029 }}$   & 0.739 ${_{ \pm0.035 }}$        & \textbf{0.739} ${_{ \pm0.030 }}$     \\
				\bottomrule
			\end{tabular}
		\end{threeparttable}
	}
\end{table*}

\subsection{Effectiveness of the Multi-Branch Encoder Block}

We also investigated the effectiveness of our proposed multi-branch encoder block, by comparing the following four models:
\begin{enumerate}
	\item Vanilla ViT model, which consisted of four sequentially connected traditional encoder blocks.
	\item Our proposed MBMD Transformer, which replaced the second and fourth traditional encoder blocks of the vanilla ViT model with the multi-branch encoder block.
	\item MBMD Transformer-1, which replaced the last traditional encoder block of the vanilla ViT model by a multi-branch encoder block.
	\item MBMD Transformer-2, which replaced the last two blocks of the vanilla ViT model by two multi-branch encoder blocks.
\end{enumerate}

\begin{figure}[htbp]\centering
	\subfigure[]{\label{fig:sen_FFN_chsz}   \includegraphics[width=\linewidth,clip]{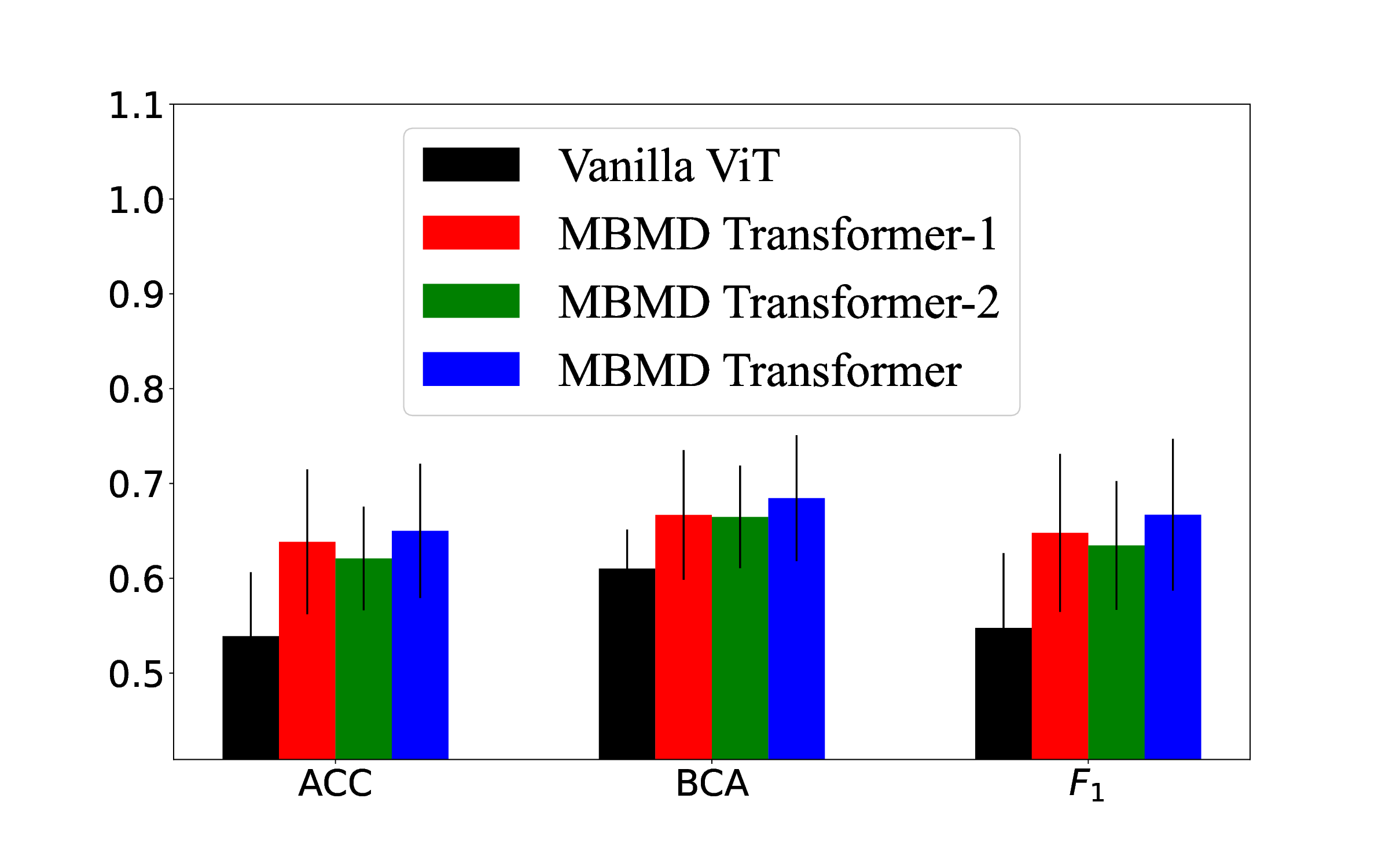}}
	\subfigure[]{\label{fig:sen_FFN_tusz}    \includegraphics[width=\linewidth,clip]{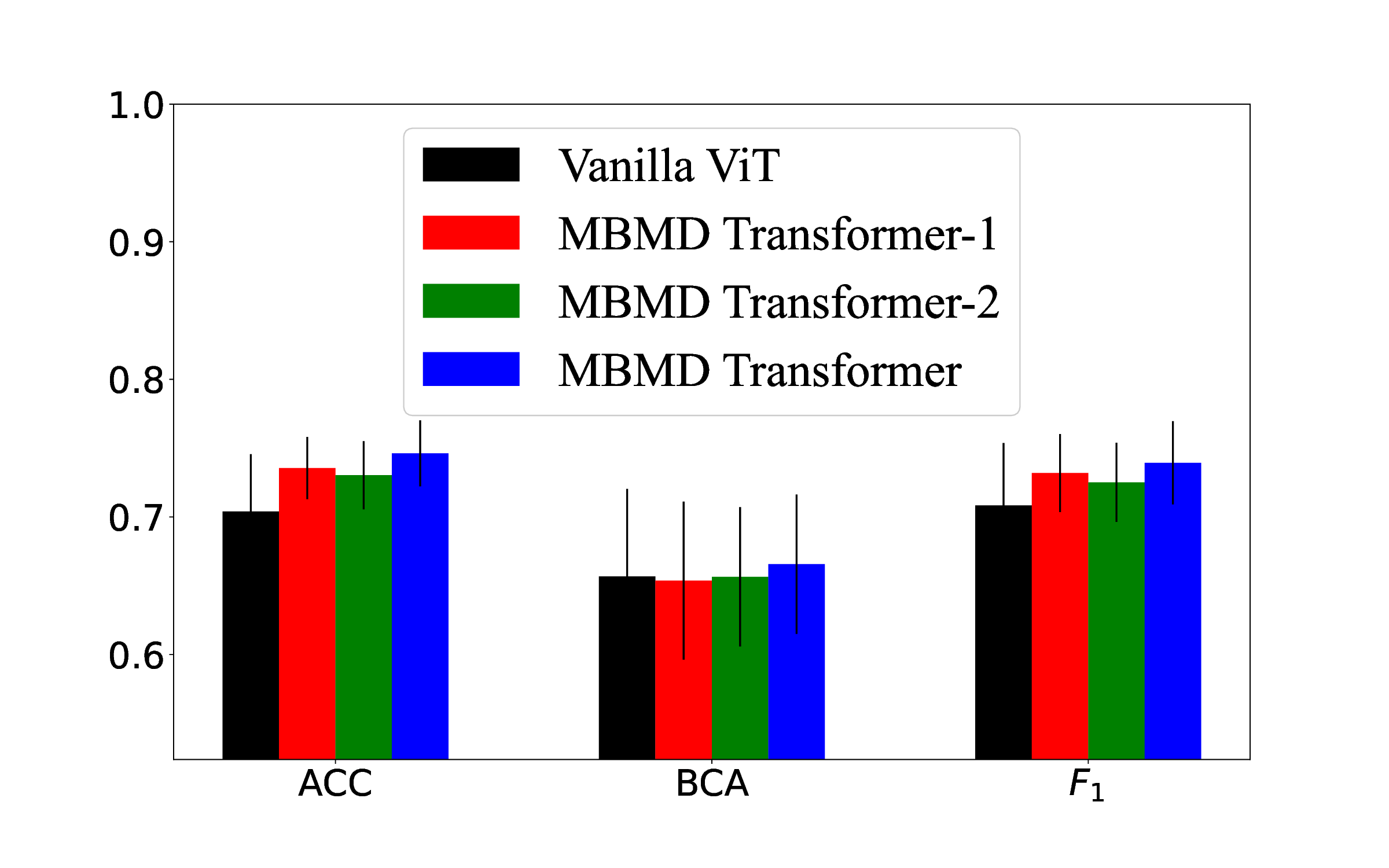}}
	\caption{Performance of the vanilla ViT and three MBMD Transformer variants on (a) CHSZ; and, (b) TUSZ.} \label{fig:sen_FFN}
\end{figure}

Fig.~\ref{fig:sen_FFN} shows the results. Our proposed MBMD Transformer achieved the best performance, but its two variants also outperformed the vanilla ViT model, suggesting the effectiveness of the multi-branch encoder block.

\subsection{Parameter Sensitivity Analysis}

This subsection studies the sensitivity of MBMD Transformer performance to its two important parameters, distillation temperature $T$ and the number of wavelets.

\begin{figure}[htbp]\centering
	\subfigure[]{\label{fig:sen_wave_chsz}   \includegraphics[width=\linewidth,clip]{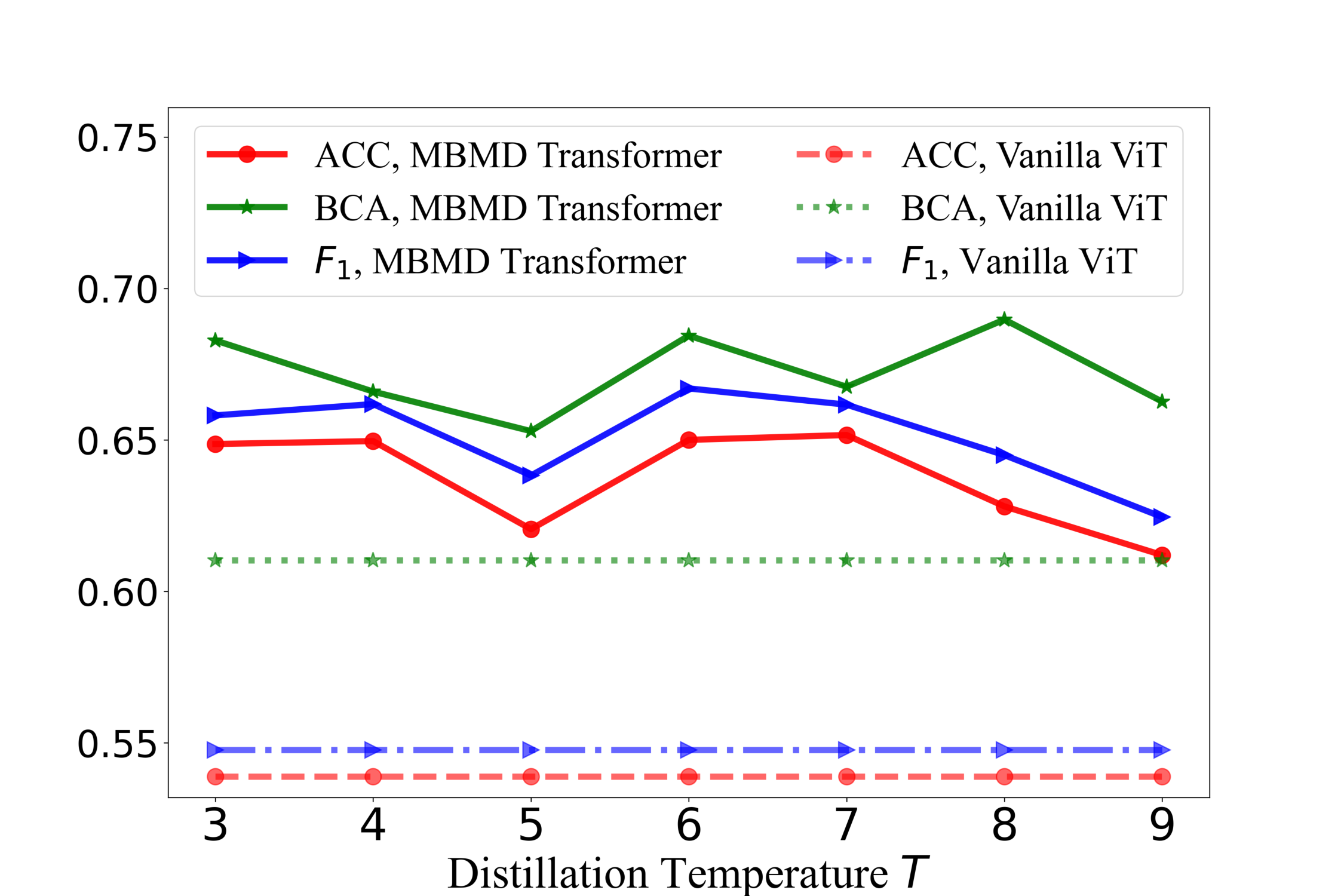}}
	\subfigure[]{\label{fig:sen_wave_tusz}    \includegraphics[width=\linewidth,clip]{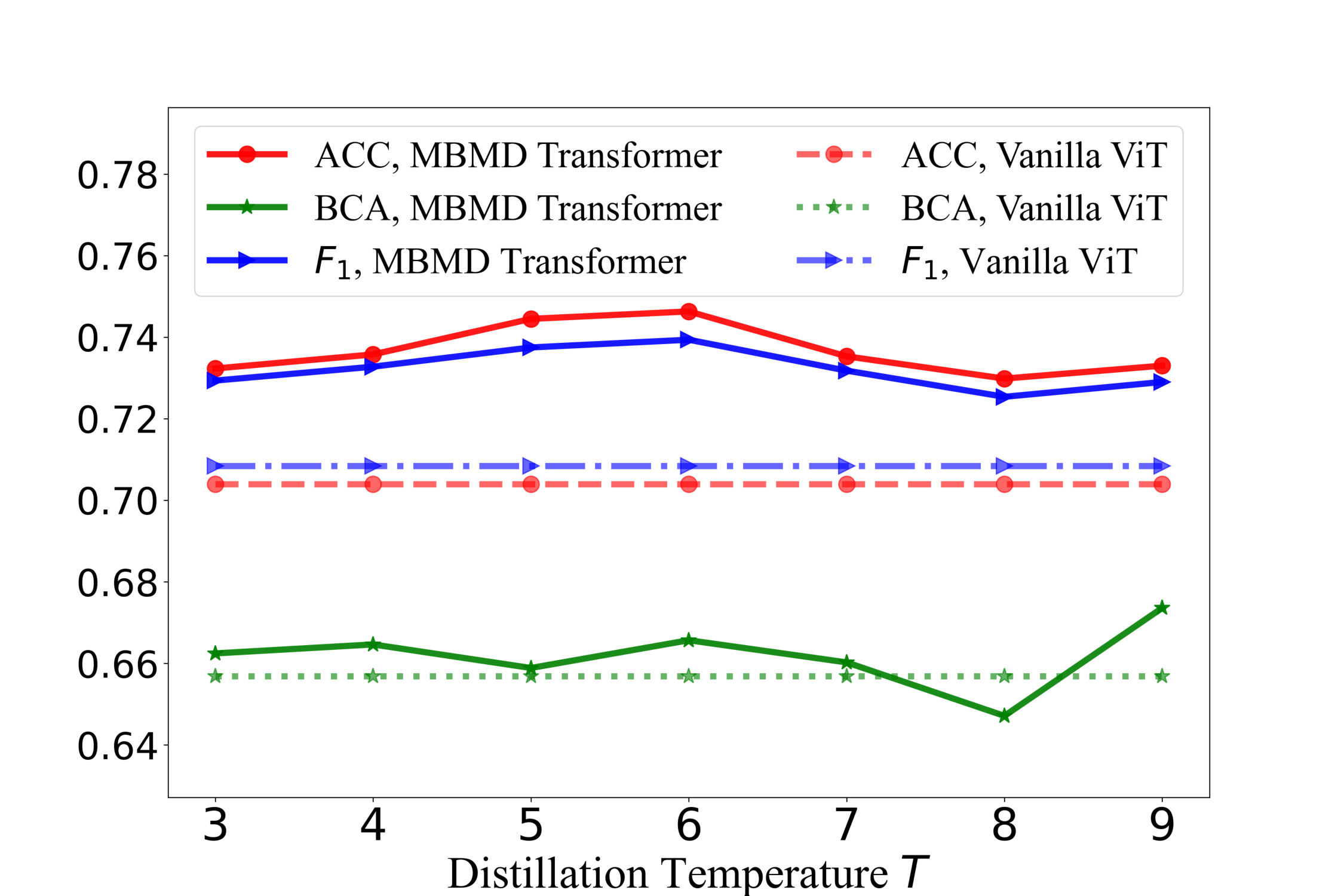}}
	\caption{Sensitivity of MBMD Transformer to the distillation temperature $T$ on (a) CHSZ; and, (b) TUSZ.} \label{fig:sen_temp}
\end{figure}

Fig.~\ref{fig:sen_temp} shows the results for $T\in[3,9]$. Generally, on both datasets, MBMD Transformer achieved higher ACC, BCA and $F_1$ scores than ViT for all $T$; however, different distillation temperatures resulted in different performance improvements. So, it is desirable to use a validation set to pick the optimal $T$.

To study the sensitivity of MBMD Transformer performance to the number of wavelet branches, we designed two MBMD Transformer variants, with two and three wavelet branches, respectively. Specifically, in the 2-branch MBMD Transformer, $\delta$, $\theta$ and $\alpha$ wavelets were merged into a single low-frequency band ($0-16$ Hz), and the remaining wavelets were merged into a single high-frequency band ($16-128$ Hz). The three frequency bands used in the 3-branch MBMD Transformer were $\delta\cup\theta$, $\alpha\cup\beta$ and $\gamma\cup others$, respectively.

\begin{figure}[htbp]\centering
	\subfigure[]{\label{fig:sen_wave_chsz}   \includegraphics[width=\linewidth,clip]{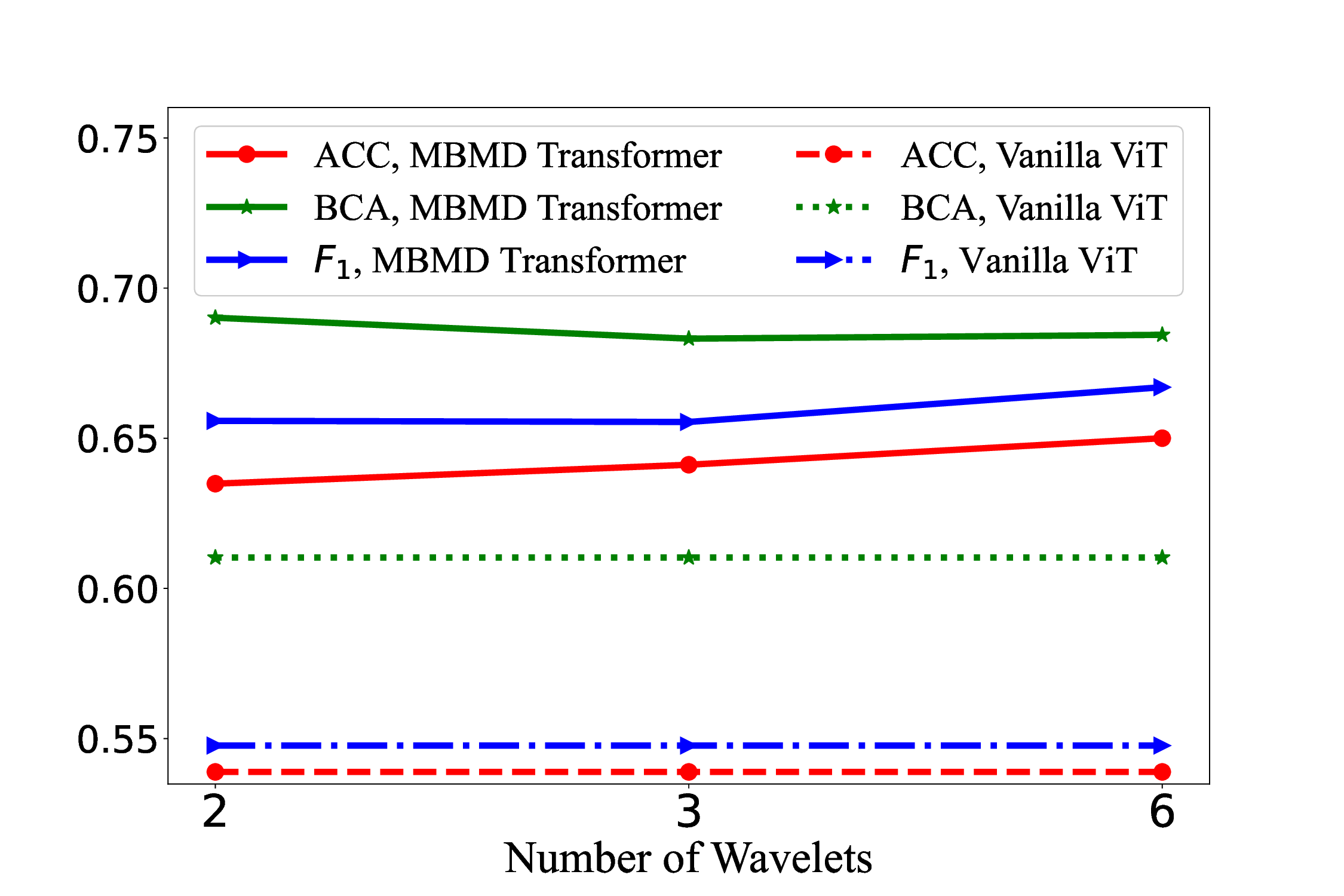}}
	\subfigure[]{\label{fig:sen_wave_tusz}    \includegraphics[width=\linewidth,clip]{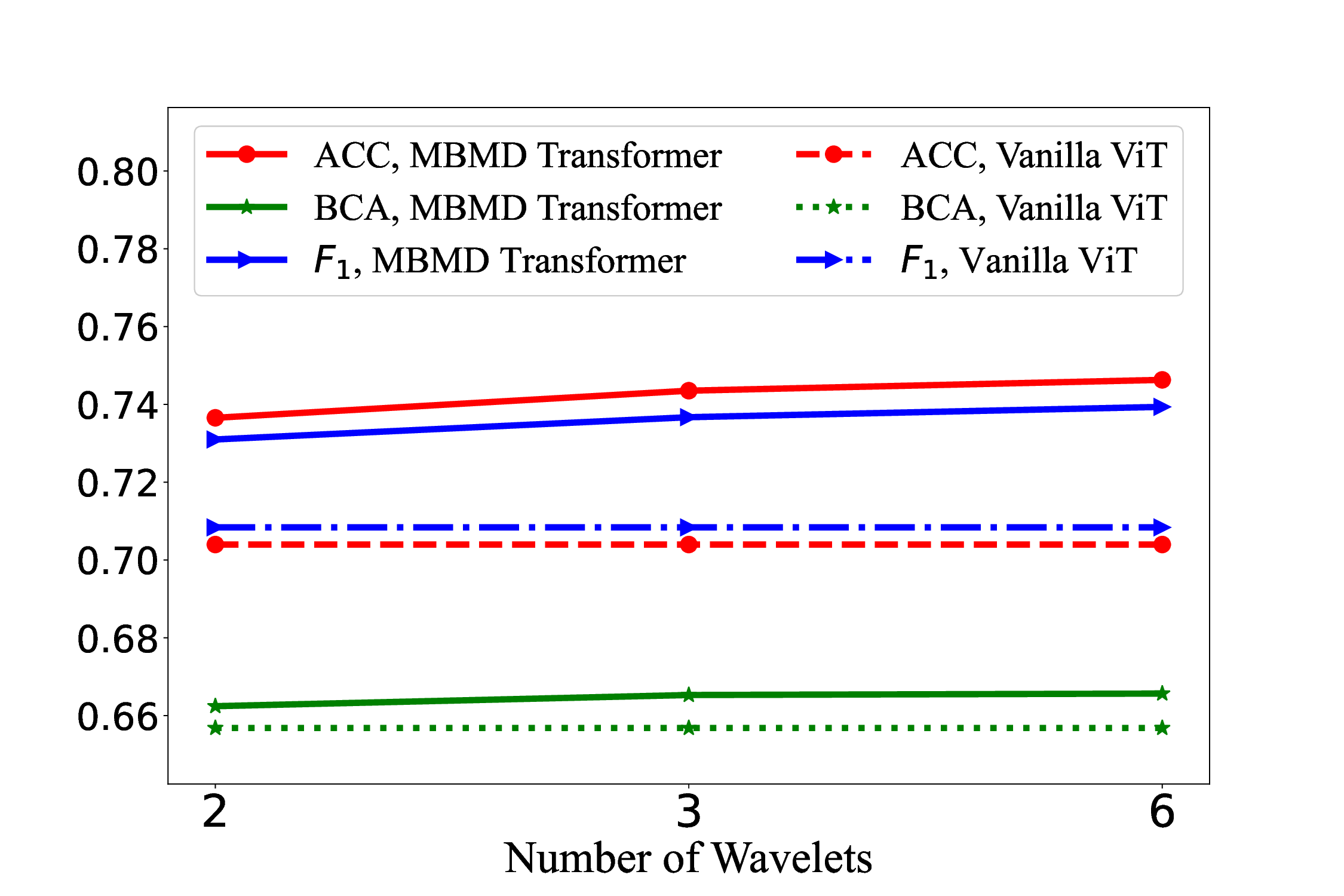}}
	\caption{Sensitivity of MBMD Transformer to the number of wavelet branches on (a) CHSZ; and, (b) TUSZ.} \label{fig:sen_waves}
\end{figure}

Fig.~\ref{fig:sen_waves} shows the results. All three MBMD Transformer variants performed similarly on both datasets, all outperforming the vanilla ViT model. In conclusion, our proposed MBMD Transformer is not sensitive to the number of wavelet branches.

\subsection{Discussion}

To validate that our proposed mutual-distillation strategy could enhance learning by ensembling and distilling the `multi-view' \cite{allen2020towards} feature sets from the auxiliary wavelet branches, we conducted the following three experiments:
\begin{enumerate}
\item \emph{Experiment 1}: Training a vanilla ViT model on the raw EEG data, and testing on the test set's raw and wavelets data. This experiment aimed to demonstrate how the single ViT trained on the raw data performs on different wavelets.
\item \emph{Experiment 2}: Training a unique vanilla ViT model for each wavelet, and testing on the corresponding wavelet of the test data. This experiment aimed to investigate the ViT learning ability on each wavelet.
\item \emph{Experiment 3}: Training our proposed MBMD Transformer, and testing on the raw data and each wavelet of the test data. This experiment aimed to validate that our proposed MBMD Transformer can achieve better performance on the raw data by utilizing latent information (features) from the wavelets.
\end{enumerate}

\begin{figure}[htbp]\centering
	\subfigure[]{\label{fig:sen_wave_chsz}   \includegraphics[width=\linewidth,clip]{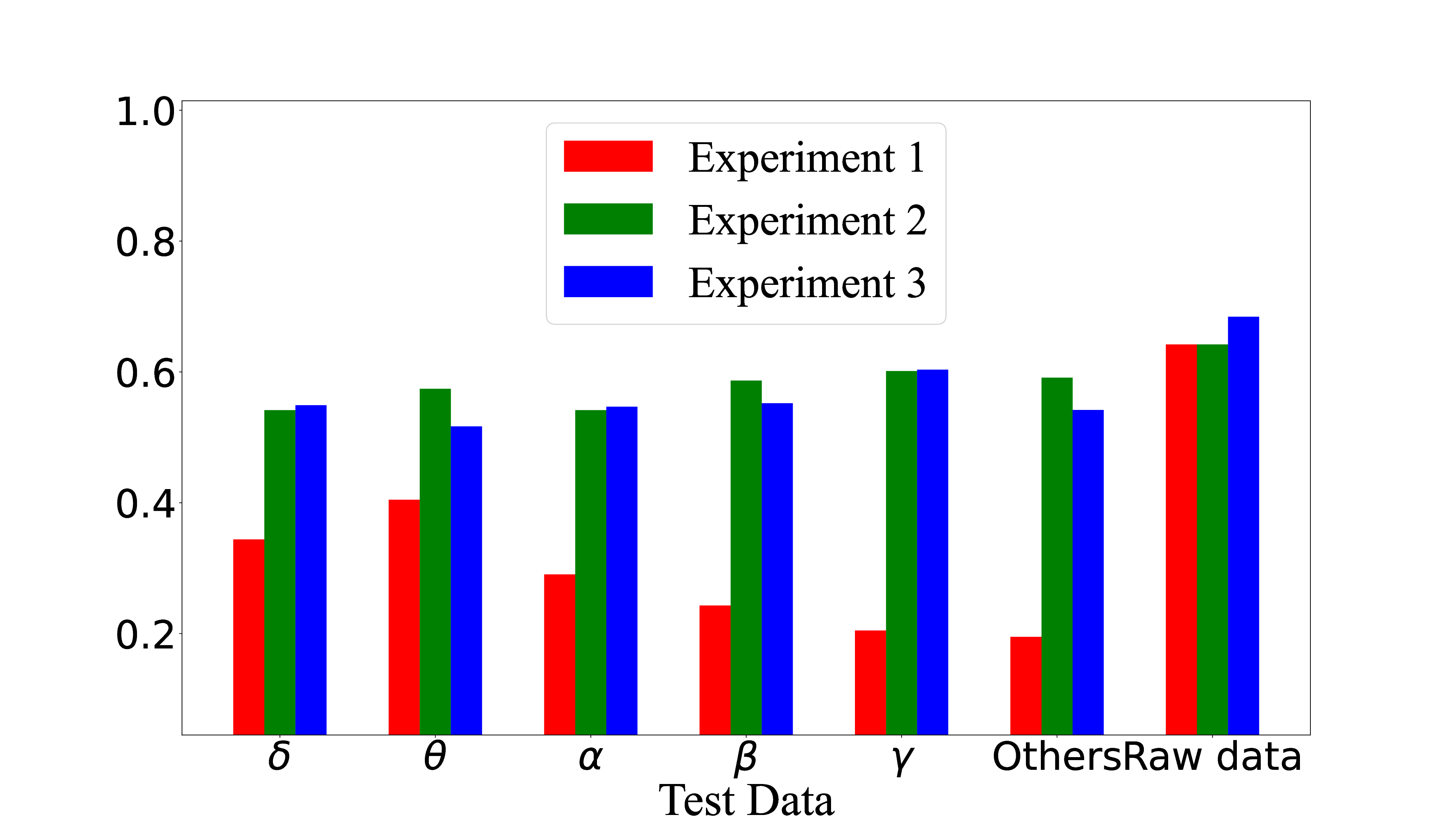}}
	\subfigure[]{\label{fig:sen_wave_tusz}    \includegraphics[width=\linewidth,clip]{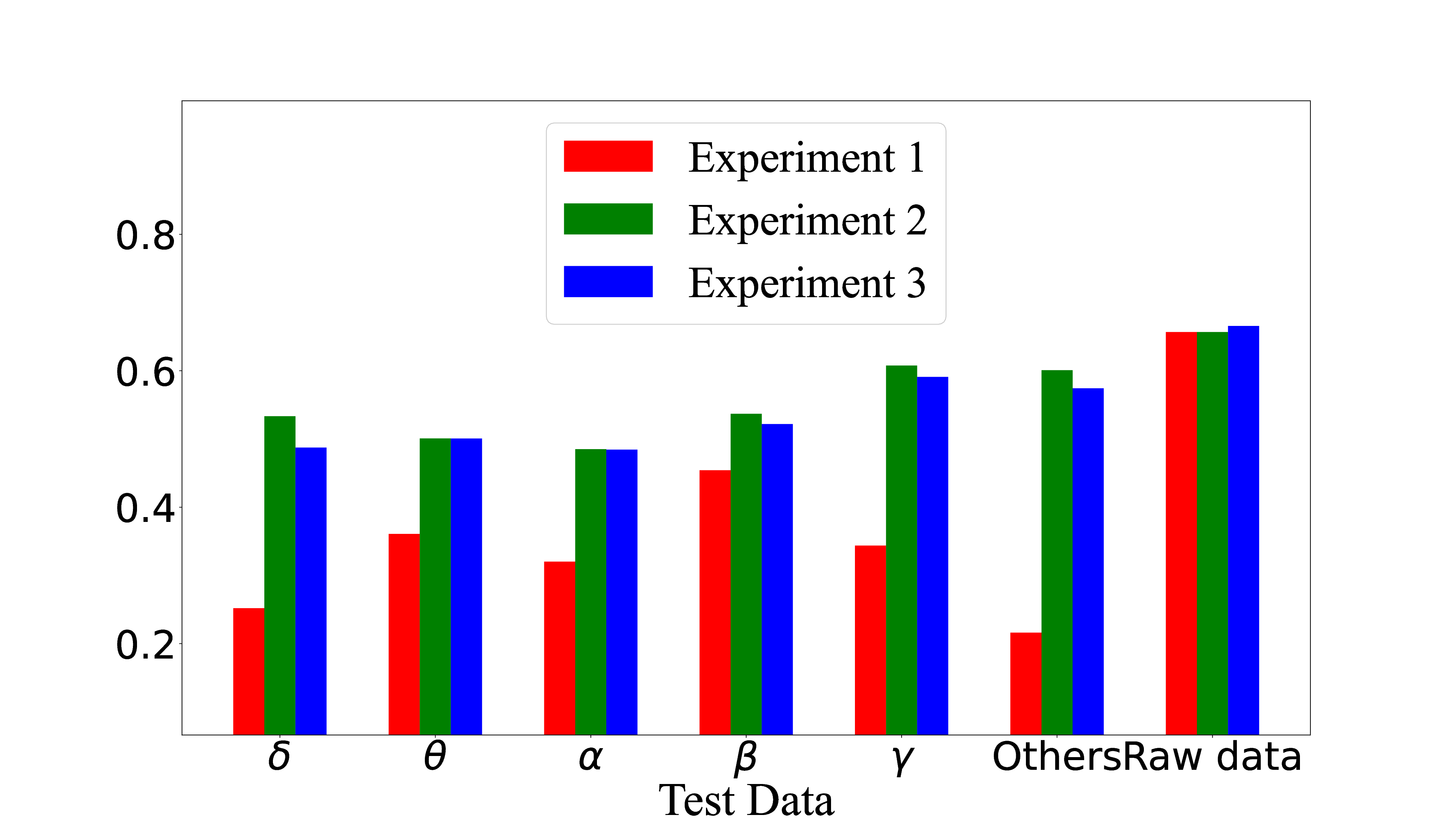}}
	\caption{BCAs on six different wavelets and the raw data on (a) CHSZ; and, (b) TUSZ.} \label{fig:analysis}
\end{figure}

Fig.~\ref{fig:analysis} shows the results. The single ViT trained in Experiment 1 had low BCAs on all six wavelets, due to the mismatching between training and test. The separate ViTs trained in Experiment 2 achieved generally the highest BCAs on the corresponding wavelets, due to perfect matching between training and test. Finally, compared with the ViTs in Experiment 2, our proposed MBMD Transformer in Experiment 3 achieved comparable or only slightly lower BCAs on the individual wavelets, but higher BCAs on the raw data, validating the benefits of utilizing latent information (features) from the wavelets.

\section{Conclusions and Future Research}

This paper has proposed MBMD Transformer for EEG-based seizure subtype classification. It replaces all even-numbered encoder blocks of the vanilla ViT model by multi-branch encoder blocks, which use mutual-distillation between the raw EEG data and its wavelet branches to transfer knowledge between them. Experiments on the public CHSZ and TUSZ datasets demonstrated that our proposed MBMD Transformer outperformed several traditional machine learning and state-of-the-art deep learning approaches in cross-subject seizure subtype classification.

The following directions will be considered in our future research:
\begin{enumerate}	
	\item Data augmentation in MBMD Transformer considered only the time-spectral characteristics of EEG signals. Other augmentation strategies, e.g., time domain and frequency domain, could also be explored.
	
	\item This paper only considered supervised training. Similar to $\pi-$model\cite{laine2016temporal}, we could also consider semi-supervised training, which minimizes the difference between the outputs of the branches and the ensemble.
	
	\item MBMD Transformer may also be applied to other frequency sensitive BCI paradigms, e.g., sleep stage classification \cite{drwuICASSP2024}.
\end{enumerate}


\end{document}